\documentclass[10pt]{article}

\usepackage{booktabs}
\usepackage{censor}
\usepackage{booktabs}
\usepackage{tabularx}
\usepackage{amsmath}
\usepackage[hidelinks]{hyperref}
\usepackage[capitalise]{cleveref}
\usepackage{todonotes}
\usepackage{inconsolata}
\usepackage{comment}
\usepackage{xcolor}
\crefformat{section}{\S#2#1#3}
\crefformat{subsection}{\S#2#1#3}
\crefformat{subsubsection}{\S#2#1#3}

\usepackage[letterpaper]{geometry}
\usepackage{hicss}
\usepackage{times}
\usepackage[none]{hyphenat}
\usepackage{url}
\usepackage{latexsym}
\usepackage{minted}
\usepackage{indentfirst}
\usepackage{graphicx}
\graphicspath{{images/}}
\usepackage[
    style=apa,
  ]{biblatex}
\title{Beyond FAIR: A Fitness Function Framework for Sustainable Research Software}

\author{Philipp Zech\\
  Department of Computer Science\\University of Innsbruck\\
  {\underline{philipp.zech@uibk.ac.at}} \\\And
  Irdin Pekaric \\
  Department of Computer Science and Information Systems\\University of Liechtenstein\\
  {\underline{irdin.pekaric@uni.li}} \\
}

\author{
Irdin Pekaric \\
  Department of Computer Science and Information Systems\\University of Liechtenstein\\
  {\underline{irdin.pekaric@uni.li}} \\\And
Philipp Zech\\
  Department of Computer Science\\University of Innsbruck\\
  {\underline{philipp.zech@uibk.ac.at}}
}

\date{}

\begin{document}
\maketitle
\begin{abstract}
\noindent    Research software sustainability is often assessed through the FAIR principles: findability, accessibility, interoperability and reusability. While important, FAIR does not cover all relevant sustainability concerns. Research software should also be environmentally responsible and secure over time. Excessive resource consumption increases environmental cost, while insecure software creates maintenance overhead, technical debt and barriers to reuse. To this end, in this paper, we extend a previously proposed fitness function framework for sustainable research software beyond FAIR. We introduce two additional sets of fitness functions: environmental functions targeting resource efficiency and execution footprint, and security functions targeting dependency health, vulnerability exposure and secure configuration. Together, these functions broaden continuous software assessment from FAIR compliance to a more complete view of sustainability. The resulting framework treats research software sustainability as a multidimensional property that includes FAIR, environmental responsibility and long-term security.
\end{abstract}

\subsubsection*{Keywords:}

Research Software, FAIR, Software Security, Environmental Sustainability, Fitness Functions

\section{Introduction}
\label{sec:introduction}

\noindent Research software is a critical output of modern science. It encodes methods, enables reproducibility and allows others to build on published results. The importance of software for contemporary scientific research~\parencite{howison2015understanding} has already been surveyed by \textcite{hannay2009scientists}. They report that 91\% of survey participants said using scientific software was important for their research. Yet sustaining it over time remains a persistent challenge. Despite its relevance, research software is not required to meet the requirements that are usually a must for other scientific methods: being peer-reviewed, being reproducible, and allowing one to build upon others' work~\parencite{jimenez2017four,morin2012shining}. Recent findings highlight this concretely, e.g., a mere 3\% of papers presented at HICSS from 2017 to 2024 have a functional and publicly available repository containing supplementary materials~\parencite{pekaric2025we}.

The FAIR principles\textemdash Findability, Accessibility, Interoperability, and Reusability\textemdash represent a fundamental framework for addressing this~\parencite{wilkinson2016fair}. While these principles were initially conceptualized for scientific data, their application has been extended to also cover research software~\parencite{lamprecht2020towards,hasselbring2020fair,barker2022introducing}. FAIR was developed not only to apply to data in the conventional sense but also for processes, workflows and tools. Availability of data and associated metadata alone cannot make research transparent, reproducible, and reusable~\parencite{radha2025advancing}.

Efforts to operationalize FAIR for research software have gained traction. \textcite{martin2024fairsoft} present FAIRsoft, a comprehensive series of quantitative indicators based on a pragmatic interpretation of the FAIR principles, providing objective indications for improving research software quality. In other work, \textcite{zech2025sustaining} proposed a complementary approach by leveraging fitness functions from evolutionary architecture~\parencite{ford2022building} as automated, continuously evaluated metrics to ensure software meets desired non-functional qualities over time. They define four fitness functions targeting findability, accessibility, interoperability, and reusability. Together, these functions form a coherent, low-overhead mechanism for evaluating FAIR compliance in research software.

But FAIR alone is not enough. Current work focuses on a narrow slice of the broader sustainability landscape. Other critical aspects, such as security, provenance, community governance, and archival persistence, are not yet represented in existing work~\parencite{martin2024fairsoft,zech2025sustaining,wilkinson2016fair}. The broader sustainability picture also includes environmental cost, e.g., facilities like LEAPS (League of European Accelerator-based Photon Sources) face challenges in preserving research data for the future while keeping in mind energy costs such as carbon footprint and limited budget~\parencite{gotz2023leaps}. These concerns apply equally to the software producing that data. On the security side, insecure software creates maintenance overhead, technical debt, and barriers to reuse that directly affect long-term viability. Research software is essential for modern scholarship but frequently encounters sustainability issues stemming from poor management methods~\parencite{carver2022survey}, and these go well beyond what FAIR currently captures.

In this paper, we extend the fitness function framework of \textcite{zech2025sustaining} beyond FAIR. We introduce two additional sets of fitness functions: environmental functions targeting resource efficiency and execution footprint, and security functions targeting dependency health, vulnerability exposure, and secure configuration \parencite{Pekaric2021Vulnerlizer, Pekaric2024AttackTreeGeneration}. Together, these functions broaden continuous software assessment from FAIR compliance to a more complete view of sustainability. The resulting framework treats research software sustainability as a multidimensional property that includes FAIR, environmental responsibility and long-term security. The framework is part of our repo\footnote{\url{https://github.com/irdin-pekaric/HICSS2027-Fitness}}.

\section{Background and Related Work}
\label{sec:background}

\noindent Research software sustainability has received growing attention over the past decade, driven by increasing recognition that scientific progress depends not only on published results but on the tools and artifacts that produce them. We briefly introduce the concept of fitness functions and then survey related work on FAIR and its operationalization for research software.

\subsection{Fitness Functions}

\noindent \textcite{ford2022building} introduce fitness functions in the context of evolutionary architectures, i.e., automated checks that continuously assess whether a software system upholds desired non-functional properties such as performance, security and maintainability. Rather than treating architectural quality as a one-time concern, fitness functions are integrated into the development life cycle and executed repeatedly, thereby flagging regressions before they become structural problems. Each function is defined as atomic, meaning it targets a single, well-scoped property and returns a clear result. \textcite{zech2025sustaining} transferred this concept to research software, proposing four fitness functions aligned with the FAIR dimensions. We build directly on that work and extend it to cover environmental and security properties not captured by FAIR.

\subsection{Related Work}

\noindent Research software sustainability is a well-documented challenge~\parencite{carver2022survey}. Scientists often lack formal software engineering training and have limited resources for quality assurance, leading to artifacts that are difficult to maintain, replicate, or reuse~\parencite{heaton2015claims,johanson2018software}.

The FAIR principles were introduced by \textcite{wilkinson2016fair} for scientific data management and have since been extended to research software~\parencite{lamprecht2020towards,barker2022introducing,hasselbring2020fair}. \textcite{barker2022introducing} present the first community effort to translate FAIR for research software, proposing a set of high-level indicators known as FAIR4RS. \textcite{lamprecht2020towards} provide an earlier formulation grounded in the life sciences domain.

Operationalizing FAIR through measurable indicators has been explored by \textcite{martin2024fairsoft},
who propose FAIRsoft, a set of quantitative indicators implemented on OpenEBench~\parencite{capella2017lessons} covering over 43,000 life sciences tools. Their results show that research software is generally findable but
barely interoperable, with significant gaps in licensing and dependency documentation. \textcite{hasselbring2020fair} further argue for open-source availability as a prerequisite for meaningful FAIR compliance.

Beyond research software, FAIR principles have been applied to machine learning (ML) workflows and models. \textcite{radha2025advancing} review tools and frameworks that support FAIR compliance across the ML life cycle, noting that availability of metadata and tools alone is insufficient without rich provenance information and standardized documentation. Beyond FAIR compliance, deploying research prototypes also requires attention to resource constraints \parencite{ave2026botlane}. \textcite{pekaric2025we} highlight the broader problem empirically, finding that only 3\% of HICSS papers from 2017 to 2024 have a functional and publicly available artifact repository, underscoring the gap between stated principles and actual practice.

\subsection{Research Gap}

\noindent Despite these efforts, existing approaches share a common limitation: they treat FAIR as the boundary of research software sustainability. Security and environmental sustainability remain largely unaddressed. A software artifact can be findable, accessible, interoperable and reusable while still relying on abandoned or vulnerable dependencies, exposing insecure configurations or consuming staggering computational resources. These properties directly affect the long-term viability of research software but fall outside what any current FAIR-aligned framework assesses. Our work fills this gap by extending the fitness function framework of \textcite{zech2025sustaining} with two additional sets of functions that make environmental and security properties measurable and automatable within the same development workflow.

\section{Methodology}
\label{sec:methodology}

\noindent Following the Design Science Research (DSR) paradigm~\parencite{Wieringa2014DesignSM} and building on the fitness function framework of \parencite{zech2025sustaining}, we extend the existing 
artifact~\parencite{pekaric2026ffrepository} to address sustainability properties beyond FAIR. The design science problem we target is:

\begin{center}
    \fbox{\parbox{0.9\columnwidth}{
    \textbf{Improve} \textit{research software sustainability} (context)\\
    \textbf{by designing} \textit{a set of fitness functions} (artifact)\\
    \textbf{that satisfy} \textit{environmental and security sustainability 
    requirements beyond FAIR} (requirement)\\
    \textbf{to deliver} \textit{sustainable research software beyond FAIR.} 
    (goal)
    }}
\end{center}

\noindent We proceed in two steps. First, we elicit extended requirements for sustainable research software beyond FAIR (cf.~\cref{ssec:requirements}). Second, we derive a set of formal, implementation-independent fitness functions from these requirements (cf.\cref{ssec:sustainability} and~\cref{ssec:security}).

\begin{table*}[h]
\small
\centering
\caption{Extended requirements for sustainable research software beyond FAIR.}
\label{tab:requirements}
\begin{tabular}{p{1.2cm} p{3.8cm} p{10.8cm}}
\toprule
\textbf{ID} & \textbf{Requirement} & \textbf{Description} \\
\midrule
    E1 & Resource Efficiency & Research software should use computational resources efficiently, including CPU, memory, storage, and network bandwidth, in order to reduce unnecessary energy consumption and infrastructure load. \\[0.4em]
    E2 & Footprint Transparency & Research software should provide transparent information about its execution footprint, such as runtime behavior, hardware requirements, and dependency overhead, to support informed and responsible use. \\[0.4em]
    E3 & Infrastructure-Aware Deployment & Research software should support deployment choices that reduce environmental burden, for example by avoiding unnecessary platform constraints, oversized dependencies, or avoidable hardware specialization. \\[0.4em]
    E4 & Recomputation Avoidance & Research software should reduce avoidable recomputation, for example through modular execution, caching, checkpointing, or clear experiment parameterization, in order to limit wasted computational effort. \\[0.6em]
    S1 & Dependency Health & Research software should rely on actively maintained and non-vulnerable dependencies in order to reduce long-term maintenance effort and operational risk. \\[0.4em]
    S2 & Secure Configuration & Research software should avoid insecure default configurations and should not expose sensitive information, excessive privileges, or unsafe communication settings. \\[0.4em]
    S3 & Integrity and Provenance & Research software should support the integrity of its artifacts and execution environments, for example through version pinning, trusted sources, and traceable build or deployment processes. \\[0.4em]
    S4 & Maintainable Security Posture & Research software should enable ongoing security maintenance by supporting vulnerability detection, dependency updates, and understandable remediation workflows over time. \\
\bottomrule
\end{tabular}
\end{table*}

\subsection{Extended Sustainability Requirements}
\label{ssec:requirements}

\noindent \textcite{zech2025sustaining} originally derived four requirements from FAIR and used them as the basis for their fitness function definitions. We follow the same approach and elicit eight additional requirements\footnote{The eight requirements were selected because they address concerns beyond FAIR, are observable through repository or execution evidence and can be acted upon by developers. They represent an initial operational set rather than an exhaustive sustainability taxonomy (broader infrastructure and hardware-level concerns remain outside the current scope).} covering two new sustainability dimensions: environmental responsibility and long-term security. Environmental requirements address the computational cost of running, deploying and maintaining research software, concerns that grow in relevance as scientific computing scales up. Security requirements, in contrast, address the exposure, integrity, and maintainability of software artifacts over time, properties that directly affect reuse and trust but are not captured by FAIR. \cref{tab:requirements} summarizes all eight requirements.

\subsection{Fitness Functions for Sustainability}
\label{ssec:sustainability}

\noindent Given the requirements E1--E4 from \cref{tab:requirements}, we define four atomic fitness functions for environmental sustainability. Following the formal style of \textcite{zech2025sustaining}, each function targets a single property, takes an artifact $A$ (and optionally a benchmark or environment) as input, and returns a Boolean result. All functions are defined using set-theoretic notation~\parencite{hausdorff2021set}.

\subsubsection{Resource Efficiency}

ensures that a research software artifact uses computational resources in a proportionate and efficient manner for a given workload. This includes, for example, CPU time, memory consumption, storage or network usage. Since efficiency cannot be assessed in the abstract, it is evaluated relative to a predefined benchmark scenario and corresponding thresholds\footnote{The thresholds used in our prototype are configurable implementation defaults (not universal sustainability limits). In practice, benchmarks and thresholds should be defined by artifact maintainers or domain communities based on representative workloads and deployment environments. These can also be calibrated against previous executions to detect regressions over time.}. Given that an artifact $A$ stays within the accepted resource bounds for a benchmark $B$, $resourceEfficiency$ returns $True$; otherwise, it returns $False$. The formal definition of $resourceEfficiency$ is as follows:
\[
\text{resourceEfficiency}: (A, B) \to \{ \text{True}, \text{False} \}
\]
\begin{tabular}{@{}p{2.5cm}p{5cm}@{}}
\toprule
\textbf{Precondition} & \( A \neq \emptyset \land B \neq \emptyset \) (The artifact and benchmark scenario are not null). \\
\textbf{Postcondition (True)} & If the artifact's measured resource consumption under the benchmark scenario remains within the defined thresholds, the function returns success: \( \text{resourceEfficiency}(A, B) = \text{True} \). \\
\textbf{Postcondition (False)} & If the artifact exceeds one or more defined resource thresholds under the benchmark scenario, the function returns failure: \( \text{resourceEfficiency}(A, B) = \text{False} \). \\ \bottomrule
\end{tabular}

\subsubsection{Footprint Transparency}

ensures that a research software artifact $A$ provides sufficient information about its execution footprint to support informed and responsible use. This may include, for example, documented runtime behavior, platform requirements, hardware constraints, or dependency overhead. Given that the required footprint information is present and sufficiently documented, $footprintTransparency$ returns $True$; otherwise, it returns $False$. The formal definition of $footprintTransparency$ is as follows:
\[
\text{footprintTransparency}: A \to \{ \text{True}, \text{False} \}
\]
\begin{tabular}{@{}p{2.2cm}p{5.3cm}@{}}
\toprule
\textbf{Precondition} & \( A \neq \emptyset \) (The artifact is not null). \\
\textbf{Postcondition (True)} & If the artifact documents the required footprint information, e.g., runtime requirements, dependency specifications, or platform constraints, the function returns success: \( \text{footprintTransparency}(A) = \text{True} \). \\
\textbf{Postcondition (False)} & If the required footprint information is missing, incomplete, or not machine- or human-readable, the function returns failure: \( \text{footprintTransparency}(A) = \text{False} \). \\ \bottomrule
\end{tabular}

\subsubsection{Infrastructure-Aware Deployment}

ensures that a research software artifact $A$ can be deployed in a way that avoids unnecessary environmental burden. This includes, for example, portability across standard environments, bounded dependency footprints, and the absence of unnecessary hardware specialization. Since these properties depend on the target environment $E$, actual implementations may vary depending on the context. Given that an artifact satisfies the encoded deployment constraints, $infrastructureAwareness$ returns $True$; otherwise, it returns $False$. The formal definition of $infrastructureAwareness$ is as follows:
\[
\text{infrastructureAwareness}: (A, E) \to \{ \text{True}, \text{False} \}
\]
\begin{tabular}{@{}p{2.2cm}p{5.3cm}@{}}
\toprule
\textbf{Precondition} & \( A \neq \emptyset \land E \neq \emptyset \) (The artifact and environment are not null). \\
\textbf{Postcondition (True)} & If the artifact can be deployed in the given environment without unnecessary platform constraints, excessive dependency overhead, or avoidable hardware specialization, the function returns success: \( \text{infrastructureAwareness}(A, E) = \text{True} \). \\
\textbf{Postcondition (False)} & If the artifact requires unnecessary infrastructure overhead or fails to satisfy the encoded deployment constraints in the given environment, the function returns failure: \( \text{infrastructureAwareness}(A, E) = \text{False} \). \\ \bottomrule
\end{tabular}

\subsubsection{Recomputation Avoidance}

ensures that a research software artifact $A$ supports mechanisms that reduce unnecessary repeated computation. Such mechanisms may include caching, checkpointing, modular execution, or explicit parameterization of experiments and workflows. Given that an artifact provides sufficient support for avoiding unnecessary recomputation, $recomputationAvoidance$ returns $True$; otherwise, it returns $False$. The formal definition of $recomputationAvoidance$ is as follows:
\[
\text{recomputationAvoidance}: A \to \{ \text{True}, \text{False} \}
\]
\begin{tabular}{@{}p{2.2cm}p{5.3cm}@{}}
\toprule
\textbf{Precondition} & \( A \neq \emptyset \) (The artifact is not null). \\
\textbf{Postcondition (True)} & If the artifact supports one or more mechanisms to reduce unnecessary recomputation, e.g., caching, checkpointing, modular execution, or parameterized reruns, the function returns success: \( \text{recomputationAvoidance}(A) = \text{True} \). \\
\textbf{Postcondition (False)} & If the artifact provides no detectable support for reducing unnecessary recomputation, the function returns failure: \( \text{recomputationAvoidance}(A) = \text{False} \). \\ \bottomrule
\end{tabular}

\subsection{Fitness Functions for Security}
\label{ssec:security}

Given the requirements S1--S4 from \cref{tab:requirements}, we define four additional atomic fitness functions targeting the security posture of research software artifacts. Analogously to the environmental functions, each function is implementation-independent and designed to be executable as part of a standard development or continuous integration workflow. 

\subsubsection{Dependency Health}

ensures that a research software artifact $A$ relies on actively maintained and non-vulnerable external components. This includes libraries, packages, frameworks, or other declared dependencies required for building or executing the artifact. Given that all relevant dependencies satisfy the encoded health constraints, $dependencyHealth$ returns $True$; otherwise, it returns $False$. The formal definition of $dependencyHealth$ is as follows:
\[
\text{dependencyHealth}: A \to \{ \text{True}, \text{False} \}
\]
\begin{tabular}{@{}p{2.2cm}p{5.3cm}@{}}
\toprule
\textbf{Precondition} & \( A \neq \emptyset \) (The artifact is not null). \\
\textbf{Postcondition (True)} & If all declared dependencies are actively maintained and no dependency violates the encoded vulnerability or maintenance constraints, the function returns success: \( \text{dependencyHealth}(A) = \text{True} \). \\
\textbf{Postcondition (False)} & If one or more declared dependencies are outdated, abandoned, or affected by known vulnerabilities beyond the accepted threshold, the function returns failure: \( \text{dependencyHealth}(A) = \text{False} \). \\ \bottomrule
\end{tabular}

\subsubsection{Secure Configuration}

ensures that a research software artifact $A$ does not expose insecure default settings or unsafe operational configurations. This may include, for example, embedded secrets, excessive privileges, unsafe communication settings, or debug-oriented defaults that increase the attack surface. Given that no relevant insecure configuration is detected, $secureConfiguration$ returns $True$; otherwise, it returns $False$. The formal definition of $secureConfiguration$ is as follows:
\[
\text{secureConfiguration}: A \to \{ \text{True}, \text{False} \}
\]
\begin{tabular}{@{}p{2.2cm}p{5.3cm}@{}}
\toprule
\textbf{Precondition} & \( A \neq \emptyset \) (The artifact is not null). \\
\textbf{Postcondition (True)} & If the artifact contains no encoded insecure defaults or unsafe configurations, the function returns success: \( \text{secureConfiguration}(A) = \text{True} \). \\
\textbf{Postcondition (False)} & If one or more insecure settings are detected, e.g., exposed secrets, unsafe permissions, plaintext communication settings, or insecure defaults, the function returns failure: \( \text{secureConfiguration}(A) = \text{False} \). \\ \bottomrule
\end{tabular}

\subsubsection{Integrity and Provenance}

ensure that a research software artifact $A$ and its execution environment can be trusted and traced over time. This may include, for example, version pinning, trusted dependency sources, verifiable build or deployment processes, and traceable provenance metadata. Given that the encoded integrity and provenance requirements are fulfilled, $integrityProvenance$ returns $True$; otherwise, it returns $False$. The formal definition of $integrityProvenance$ is as follows:
\[
\text{integrityProvenance}: A \to \{ \text{True}, \text{False} \}
\]
\begin{tabular}{@{}p{2.2cm}p{5.3cm}@{}}
\toprule
\textbf{Precondition} & \( A \neq \emptyset \) (The artifact is not null). \\
\textbf{Postcondition (True)} & If the artifact supports integrity and provenance through mechanisms such as version pinning, trusted sources, or traceable build and deployment processes, the function returns success: \( \text{integrityProvenance}(A) = \text{True} \). \\
\textbf{Postcondition (False)} & If integrity or provenance guarantees are missing, incomplete, or unverifiable, the function returns failure: \( \text{integrityProvenance}(A) = \text{False} \). \\ \bottomrule
\end{tabular}

\subsubsection{Maintainable Security Posture}

ensures that a research software artifact $A$ supports ongoing security maintenance over time. This includes, for example, mechanisms for vulnerability detection, dependency updates, security policies, or understandable remediation workflows. Since long-term maintainability cannot be assessed directly, implementations of $maintainableSecurity$ rely on observable evidence that the artifact can be monitored and updated with reasonable effort. Given that sufficient maintenance evidence is present, $maintainableSecurity$ returns $True$; otherwise, it returns $False$. The formal definition of $maintainableSecurity$ is as follows:
\[
\text{maintainableSecurity}: A \to \{ \text{True}, \text{False} \}
\]
\begin{tabular}{@{}p{2.2cm}p{5.3cm}@{}}
\toprule
\textbf{Precondition} & \( A \neq \emptyset \) (The artifact is not null). \\
\textbf{Postcondition (True)} & If the artifact provides sufficient evidence of ongoing security maintenance, e.g., automated scanning, update workflows, security policies, or clear remediation support, the function returns success: \( \text{maintainableSecurity}(A) = \text{True} \). \\
\textbf{Postcondition (False)} & If the artifact provides no sufficient evidence that security issues can be detected and remediated with reasonable effort over time, the function returns failure: \( \text{maintainableSecurity}(A) = \text{False} \). \\ \bottomrule
\end{tabular}

The proposed fitness functions\footnote{Although some functions use related evidence, they assess different properties: \textit{Footprint Transparency} evaluates if execution requirements are documented for users, \textit{Infrastructure Awarenes}s evaluates if the software imposes avoidable deployment or hardware constraints, \textit{Dependency Health} assesses the current condition of dependencies and \textit{Maintainable Security} assesses the processes available to detect and remediate future security issues.} operationalize sustainability through observable repository and execution-level indicators and not by providing complete measurements of environmental impact or security. Their outcomes should be interpreted as lightweight automated evidence for the corresponding sustainability properties.

\section{Results}
\label{sec:results}

\noindent In this section, we report the implementation and evaluation of the extended fitness function framework. \textcite{zech2025sustaining} already demonstrated that research software sustainability can be translated into lightweight, modular and executable checks that inspect repository metadata, source-code structure and execution environments.

The new prototype follows the same logic. Each  function is implemented as an independent checker that takes artifact metadata as input, retrieves or inspects the corresponding repository, applies a set of verification steps and returns a Boolean result together with diagnostic evidence. This keeps the implementation aligned with the modular structure of the original FAIR Artifact Checker \parencite{zech2025sustaining}.

```latex
\begin{table}[t]
\centering
\caption{Comparative results across artifacts for E1--E4 and S1--S4 (Table~1).}
\label{tab:comparative_results}
\scriptsize
\setlength{\tabcolsep}{2.2pt}
\begin{tabular}{lcccccccc}
\hline
\textbf{Artifact} &
\textbf{E1} & \textbf{E2} & \textbf{E3} & \textbf{E4} &
\textbf{S1} & \textbf{S2} & \textbf{S3} & \textbf{S4} \\
\hline
HICSS        & P & P & P & F & P & P & P & F \\
TCN Dataset  & N/E & N/E & N/E & N/E & N/E & N/E & N/E & N/E \\
Python Alg.  & P & P & F & P & P & F & P & P \\
Scikit-learn & P & P & P & P & P & F & P & P \\
Pandas       & P & P & P & P & P & F & P & P \\
NumPy        & P & P & P & P & P & F & P & P \\
Matplotlib   & P & P & P & P & P & F & P & P \\
Scrapy       & P & P & F & P & F & F & P & P \\
Flask        & P & P & F & P & P & F & P & P \\
TensorFlow   & P & P & F & P & F & F & F & P \\
\hline
\end{tabular}

\vspace{1mm}
\raggedright
\scriptsize
\textit{P = passed; F = failed; N/E = not evaluated because the repository was unavailable.}
\end{table}
```

The evaluation is based on the artifacts listed in the metadata file, including the original HICSS artifact repository \parencite{pekaric2025we} and several widely used Python-based repositories such as scikit-learn, pandas, NumPy, Matplotlib, Scrapy, Flask, and TensorFlow. The artifact metadata contains titles, repository URIs, and descriptions for each evaluated artifact. Tab.~\ref{tab:comparative_results} summarizes the
results of all eight fitness functions across the evaluated
artifacts. We subsequently use the HICSS artifact as a
running example to illustrate the diagnostic evidence
produced by individual checkers.

To sanity-check the automated results, we manually inspected one passing and, where available, one failing outcome for each fitness function against the corresponding repository evidence. The inspected outcomes were consistent with the indicators reported by the checkers. This provides an initial evidence that the implementations detect the repository-level properties they are intended to operationalize (this does not establish that these indicators fully capture the broader sustainability constructs).

\subsection{Environmental Fitness Functions}

\noindent In the following, we describe implementation and results  of the environmental fitness functions. The \texttt{ResourceEfficiencyChecker} 
evaluates if an artifact stays within predefined resource thresholds during a benchmarked execution. For the HICSS artifact, the checker resulted in a small repository footprint of 1.58 MB, 48 files and three dependencies. The execution finished successfully in 0.42 seconds with a peak memory consumption of 10.79 MB, which resulted in passing the resource-efficiency check under the configured benchmark scenario. 

This result indicates that the artifact can be executed with limited computational overhead in the tested environment. However, the result should not be interpreted as a universal claim about the artifact’s environmental cost. Resource efficiency is necessarily benchmark-dependent: a lightweight benchmark confirms that the artifact can be executed within bounded resource limits, but more demanding scientific workloads may require additional measurements. Nevertheless, the checker provides a practical first step for making computational cost visible and testable.

The \texttt{FootprintTransparencyChecker} evaluates if an artifact provides sufficient information about its execution footprint. For the HICSS artifact, the checker found a \texttt{README.md} file and a \texttt{requirements.txt} file, which provide basic human-readable documentation and machine-readable dependency information. However, the checker did not detect stronger forms of footprint documentation (e.g., container specifications, explicit hardware requirements or continuous-integration evidence). This resulted in an artifact passing the check, but only at a basic level. Such result shows the value of distinguishing between minimal and stronger evidence. The artifact provides enough information for a user to understand and install the software, but it does not yet provide a fully transparent execution profile. Thus, adding information regarding expected runtime, memory requirements, supported platforms or containerized execution would significantly improve the artifact’s footprint transparency.

\begin{figure}[htbp]
    \centering
    \includegraphics[width=1\columnwidth]{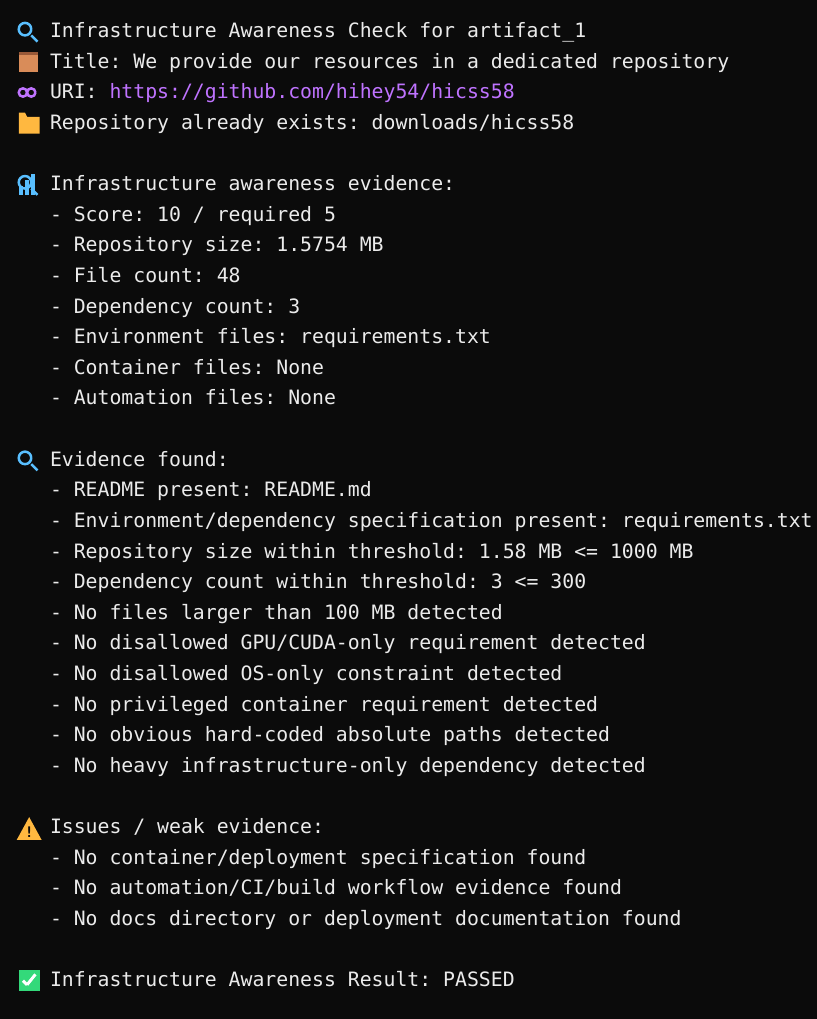}
    \caption{Output of the \texttt{InfrastructureAwarenessChecker}.}
    \label{fig:infrastructure}
\end{figure}

The \texttt{InfrastructureAwarenessChecker} (see Fig. \ref{fig:infrastructure}) 
evaluates if an artifact can be deployed without unnecessary infrastructure burden. For the HICSS artifact, the checker indicated a small repository size, a low number of dependencies, no large files, and no obvious GPU, CUDA, OS-specific, privileged-container or hard-coded path constraints. The artifact therefore passed the infrastructure-awareness check. At the same time, the checker also reported missing evidence for container or deployment specifications, CI workflows and explicit deployment documentation. This means that the artifact does not appear to impose not required infrastructure constraints, but it also does not provide real guidance for reproducible deployment. The result is positive but can be improved --- the repository is lightweight and portable, but its deployment process could be documented more explicitly.

The \texttt{RecomputationAvoidanceChecker} evaluates if an artifact provides mechanisms to reduce unnecessary repeated computation. For the HICSS artifact, the checker detected some basic source-code structure and workflow evidence, but not enough to pass the configured threshold. In particular, it did not find sufficient evidence of caching, checkpointing, modular experiment execution, configuration files, notebooks or any real parameterized reruns. The resulted in a failed check, which highlights a sustainability gap that is not captured by the original FAIR functions. A repository may be findable, accessible, and reusable, while still requiring users to rerun computations unnecessarily. For the evaluated artifact, recomputation avoidance could be improved by adding clear experiment parameters, modular execution scripts, intermediate result caching or checkpointing support.

\subsection{Security Fitness Functions}

\begin{figure}[htbp]
    \centering
    \includegraphics[width=0.85\columnwidth]{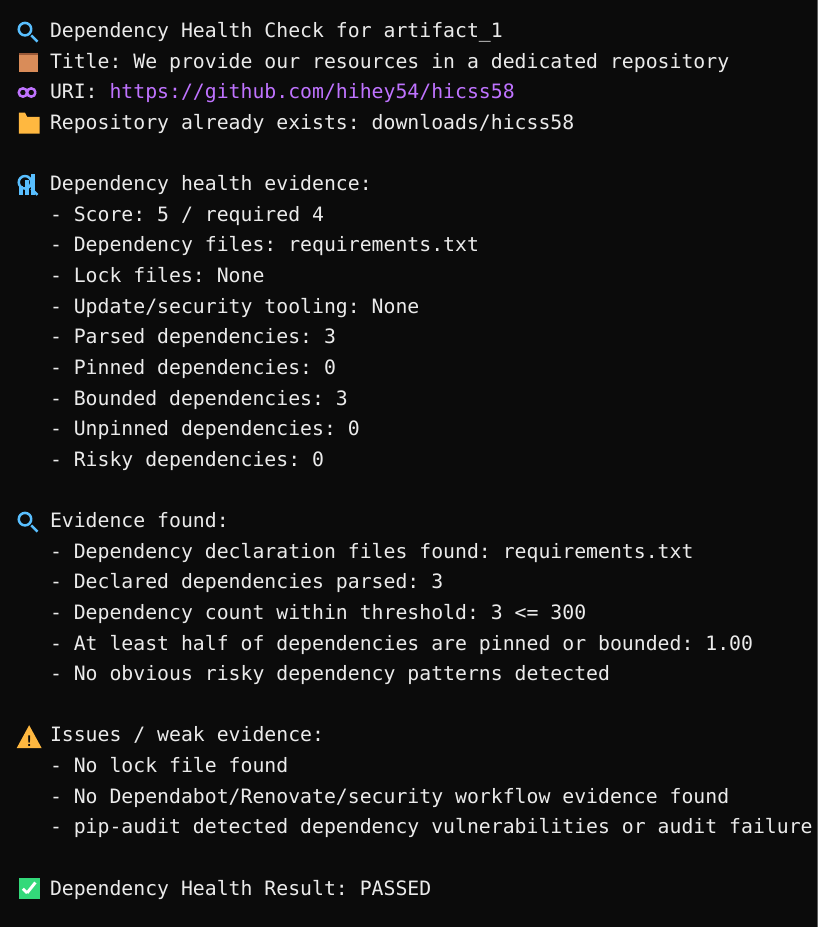}
    \caption{Output of the \texttt{DependencyHealthChecker}.}
    \label{fig:dependency}
\end{figure}

\noindent The second set of functions evaluates long-term security.
 The \texttt{DependencyHealthChecker} (see Fig. \ref{fig:dependency}) checks if an artifact uses declared, bounded, pinned or manageable dependencies. Vulnerability exposure is assessed using pip-audit, which checks declared Python dependencies against known vulnerability records. For the HICSS artifact, the checker found three declared dependencies in \texttt{requirements.txt}. These dependencies were bounded, and the checker did not detect risky patterns, which resulted in artifact passing the dependency-health check. This result shows that the repository provides a manageable dependency surface. However, dependency health is not a static property. Even when dependencies are properly declared, they may become outdated or vulnerable. This means that dependency-health checks are most useful when executed over time (e.g, as part of a continuous-integration or periodic maintenance).

The \texttt{SecureConfigurationChecker} checks if an artifact exposes insecure defaults or unsafe operational patterns. For the HICSS artifact, the checker scanned the repository and did not detect hard-coded secrets, exposed API keys, unsafe debug defaults, plaintext communication settings, disabled SSL/TLS verification, privileged container settings or committed sensitive files. These issues are detected through lightweight static pattern matching across source and configuration files (findings in examples, documentation, or test code may represent false positives). It also detected protective \texttt{.gitignore} patterns excluding \texttt{.env} files. The artifact therefore passed the secure-configuration check, wherein the result indicated that the evaluated artifact does not show obvious insecure configuration patterns. However, the checker should be understood as a diagnostic mechanism and not a complete security audit. Its purpose is to make common security risks visible early. This is specifically the case for research repositories where systematic security review is often not part of the development process.

The \texttt{IntegrityProvenanceChecker} evaluates if an artifact supports traceability and verifiability over time. For the HICSS artifact, the checker found Git metadata, a traceable commit hash, a README file, a license file, a dependency file and version-pinning evidence. Based on these indicators, the artifact passed the integrity and provenance check. The checker also reported missing stronger provenance evidence, such as \texttt{CITATION.cff}, lock files, container files, checksum files and explicit release metadata. This result shows that the artifact provides basic provenance and traceability, but could still be improved by adding machine-actionable citation metadata and stronger mechanisms for verifying exact execution environments.

The \texttt{MaintainableSecurityChecker}  
checks if an artifact provides evidence that security issues can be detected, reported and remediated over time. For the HICSS artifact, the checker found a dependency file, but did not find a security policy, dependency-update automation, security workflow, CI workflow, test directory, issue templates, governance files, release metadata, lock files or static-analysis tooling. The artifact therefore failed the maintainable-security check. This is very important because it reveals a limitation that is invisible from a FAIR-only perspective since the artifact does not provide enough evidence that security maintenance can be performed systematically over time. Adding a security policy, automated dependency updates, static analysis and basic CI checks would help improve the long-term security posture. Implementation details including evaluation of additional repositories can be found in our own repository\footnote{\url{https://anonymous.4open.science/r/HICSS2027-Fitness-E680}}.

\section{Discussion}
\label{sec:discussion}

\noindent Our implementation of environmental and security-oriented fitness functions shows that research software sustainability is not limited only to FAIR, but can still be treated as a measurable and actionable property. We make environmental responsibility and long-term security testable through minimal scripts by extending the original modular FAIR-oriented framework. The checks can be executed locally by researchers without advanced infrastructure, which is important in academic settings wherein centralized quality assurance and long-term maintenance support are often unavailable. 

Our results indicate that small omissions, such as missing caching, checkpointing, dependency-update automation, security policies, static analysis or CI workflows, can reduce a project’s long-term utility. By coding such expectations into executable checks, the framework pinpoints these issues early, before they become barriers to reuse, replication or maintenance. Regardless that our implementation targets Python-based GitHub repositories, the underlying idea and principles are language-agnostic and can be adapted to other ecosystems. The separation of concerns and ``one function per sustainability property'' supports future extension. Running these checks regularly, either in CI pipelines or manually, helps researchers detect sustainability issues early and shows that the framework can be applied to real research software artifacts.

\subsection{Sustainability Implications}

\noindent From a sustainability perspective, the extended framework extends sustainable research software beyond artifact availability and reuse. FAIR remains essential, but long-term sustainability also depends on environmental and security properties. A repository that is easy to find and download can still be costly to execute repeatedly, difficult to deploy responsibly or hard to maintain securely. With environmental fitness functions, the framework makes resource consumption, execution footprint, infrastructure requirements and recomputation avoidance visible, which helps users execute artifacts responsibly and encourages developers to document requirements, support modular execution and reduce unnecessary computation.

The framework also strengthens trust by treating security as part of sustainability. The security functions address dependency health, secure configuration, integrity/provenance and maintainable security, which are important for long-term reuse. Due to simplicity of the framework and by relying on measurable and automatable properties, the approach becomes practical for everyday research use. Researchers can execute the checks locally, interpret the diagnostic evidence and improve their repositories. This can be done without centralized quality-assurance infrastructure, which supports sustainability and provides a basis for more systematic artifact quality checks by researchers.

\subsection{Limitations and Future Work} 

\noindent Our approach offers a practical way to assess sustainability beyond FAIR principles only. However, it has several limitations. The current implementation targets Python-based GitHub repositories, while the fitness-function definitions are implementation-independent. This extends the prototype to other languages or platforms requires ecosystem-specific adaptations for dependency analysis, vulnerability checking and execution. Some checks are also heuristic: secure-configuration findings may point to patterns in documentation, tests or examples and not to actual vulnerabilities, while ``not identifying'' something does not prove that a repository is indeed secure. Thus, the framework should be seen as a lightweight first layer of automated evidence, not a replacement for manual review or security auditing \parencite{Pekaric2022SensorSpoofing,Pekaric2024AttackTreeGeneration}.

Further limitations concern benchmark dependency and binary reporting. Resource efficiency must be evaluated relative to a workload and threshold and a lightweight benchmark may not reflect realistic scientific use. Binary outcomes support simple automated CI decisions but necessarily compress differences in severity. For example, minor and extensive non-compliance may produce the same result. The accompanying diagnostic evidence therefore provides additional context, while graded scoring or maturity levels remain future work. Computational overhead also varies across functions. Most rely on lightweight repository inspection, wherein resource efficiency may increase build time because it executes a benchmark workload. More expensive checks can therefore be scheduled periodically rather than on every commit. Future work will therefore focus on configurable benchmarks, domain-specific thresholds, maturity levels, broader ecosystem support, reusable predefined fitness functions and integration into artifact-evaluation, journal, and institutional workflows. In addition, to lower the application barrier, we also plan to investigate domain-specific languages for modeling such fitness checks and subsequently generate executable checks from such fitness models~\parencite{wkasowski2023domain}. Model-based toolchains such as SAFT-GT \parencite{pekaric2026bridging} offer a precedent for automating security analysis through model transformations.

\section{Conclusion}
\label{sec:conclusion}

\noindent We extend an existing fitness function framework beyond FAIR by adding eight atomic checks for environmental and security sustainability. These functions assess resource efficiency, footprint transparency, infrastructure-aware deployment, recomputation avoidance, dependency health, secure configuration, integrity/provenance and maintainable security posture. Our results show that FAIR-compliant artifacts may still lack support for efficient repeated computation or systematic security maintenance. Thus, sustainable research software should not only be findable, accessible, interoperable and reusable, but also environmentally responsible, transparent in execution, deployable with limited infrastructure burden and maintainable from a security perspective.

\renewcommand*{\bibfont}{\small}
\printbibliography

\end{document}